\documentclass[10pt,conference]{IEEEtran}
\usepackage{graphicx}
\usepackage{amsfonts,amssymb,amsmath,alltt,amsthm}
\usepackage{url}
\usepackage{color}

\newenvironment{ttbox}{\begin{alltt}\ttbraces\small\tt}%
                      {\end{alltt}}
\def\ttbraces{\let\.=\nobreak\chardef\{=`\{\chardef\}=`\}\chardef\|=`\\}

\newcommand\ttand{\mbox{{$\land$}}}

\newcommand\ttfun{\mbox{{$\Rightarrow$}}}

\newcommand\ttimp{\mbox{{$\longrightarrow$}}}
\newcommand\ttequiv{\mbox{{$\equiv$}}}
\newcommand\ttexists{\mbox{{$\exists$}}}
\newcommand\ttforall{\mbox{{$\forall$}}}

\newcommand\ttneq{\mbox{{$\neq$}}}
\newcommand\ttin{\mbox{{$\in$}}}

\newcommand\ttImp{\mbox{{$\Longrightarrow$}}}
\newcommand\ttlam{\mbox{\( \lambda \)}}
\newcommand\tttimes{\mbox{\( \times \)}}
\newcommand\ttsigma{\mbox{{$\sigma$}}}

\newcommand{\ttcalN}[1]{\mbox{{${\mathcal{N}}_{\texttt{#1}}$}}} 
\newcommand\ttattand[1]{\mbox{{$\oplus_{\wedge}^{#1}$}}}
\newcommand\ttattor[1]{\mbox{{$\oplus_{\vee}^{#1}$}}}

\newcommand\ttrelI{\mbox{{$\to$}}}
\newcommand\ttrelIstar{\mbox{{$\to^*$}}}

\newcommand\ttsubseteq{\mbox{{$\subseteq$}}}
\newcommand\ttsupseteq{\mbox{{$\supseteq$}}}

\newcommand\ttvdash{\mbox{{$\vdash$}}}

\begin{document}

%% Title information
\title{Higher Order Model Checking in Isabelle for Human Centric Infrastructure Security}
%A Formal Development Cycle for Security Engineering in Isabelle}         %% [Short Title] is optional;
                                        %% when present, will be used in
                                        %% header instead of Full Title.
%\titlenote{with title note}             %% \titlenote is optional;
                                        %% can be repeated if necessary;
                                        %% contents suppressed with 'anonymous'
%\subtitle{Subtitle}                     %% \subtitle is optional
%\subtitlenote{with subtitle note}       %% \subtitlenote is optional;
                                        %% can be repeated if necessary;
                                        %% contents suppressed with 'anonymous'

%% Author information
%% Contents and number of authors suppressed with 'anonymous'.
%% Each author should be introduced by \author, followed by
%% \authornote (optional), \orcid (optional), \affiliation, and
%% \email.
%% An author may have multiple affiliations and/or emails; repeat the
%% appropriate command.
%% Many elements are not rendered, but should be provided for metadata
%% extraction tools.

\author{\IEEEauthorblockN{Florian Kamm\"uller}
\IEEEauthorblockA{Middlesex University London, UK\\
f.kammueller@mdx.ac.uk}}

%% Author with single affiliation.
%\author{Florian Kamm\"uller}
%\authornote{}          %% \authornote is optional;
                                        %% can be repeated if necessary
%\orcid{0000-0001-5839-5488}             %% \orcid is optional
%\affiliation{
%  \position{Position1}
%  \department{Department1}              %% \department is recommended
%  \institution{Middlesex University London and TU Berlin}            %% \institution is required
%  \streetaddress{Street1 Address}
%  \city{London and Berlin}
%  \state{State1}
%  \postcode{Post-Code1}
%  \country{UK and Germany}                    %% \country is recommended
%}
%\email{f.kammueller@mdx.ac.uk}          %% \email is recommended
%\acmConference[arxive]{ }{2020}{ }
%\authornotemark{ }
%% Abstract
%% Note: \begin{abstract}...\end{abstract} environment must come
%% before \maketitle command
%% \maketitle
%% Note: \maketitle command must come after title commands, author
%% commands, abstract environment, Computing Classification System
%% environment and commands, and keywords command.
\maketitle

\begin{abstract}
  In this paper we present an efficient approach to implementing model checking in the Higher Order Logic (HOL)
  of Isabelle. This is a non-trivial task since model checking is restricted to finite state sets. By restricting
  our scope to considering security attacks, we achieve an efficient executable specification of a model checking
  algorithm for attack trees. We provide the existing background, the necessary theory and illustrate its application.
  Theory and application are fully formalized in Isabelle thus providing an executable model checking algorithm.
\end{abstract}

%% 2012 ACM Computing Classification System (CSS) concepts
%% Generate at 'http://dl.acm.org/ccs/ccs.cfm'.
%\begin{CCSXML}
%<ccs2012>
%<concept>
%<concept_id>10011007.10011006.10011008</concept_id>
%<concept_desc>Software and its engineering~General programming languages</concept_desc>
%<concept_significance>500</concept_significance>
%</concept>
%<concept>
%<concept_id>10003456.10003457.10003521.10003525</concept_id>
%<concept_desc>Social and professional topics~History of programming languages</concept_desc>
%<concept_significance>300</concept_significance>
%</concept>
%</ccs2012>
%\end{CCSXML}

%\ccsdesc[500]{Software and its engineering~General programming languages}
%\ccsdesc[300]{Social and professional topics~History of programming languages}
%% End of generated code

%% Keywords
%% comma separated list
%\keywords{Isabelle theorem proving, Security and Privacy, IoT}  %% \keywords are mandatory in final camera-ready submission

\section{Introduction}
\label{sec:intro}
Model checking is an automated verification technique for checking properties on a system model. It is
a natural requirement for verification in system engineering as are testing, simulation and proof. Referring
to testing, model checking can be seen as a complete test of all possible system execution paths. Clearly model
checking is a complex task, in fact, it is exponential in the size of the state space. Consequently, model checking
can only work for finite state sets, otherwise it does not terminate. Even for finite but rather complex state
sets, it becomes quickly infeasible. Therefore, finding suitable abstractions of systems is an essential first step
when applying model checking as a verification method.

Higher Order Logic (HOL) based verification, like specification and proof in Isabelle, has no such limits. In fact,
infinity can be expressed there as a concept and can be reasoned with and about. ``Reasoning with'' means using a
mathematical concept, for example infinity, in specification and proof while ``reasoning about'' means that the
concept is the subject of specification and proof, for example, we prove relations between different forms of infinity.
%The expressiveness of Isabelle/HOL allows building such such logical models.
The expressiveness of Isabelle/HOL allows combining meta-logical reasoning with reasoning in object-logics.
On the other side, the verification in Isabelle/HOL is thus undecidable. Therefore, proving in Isabelle/HOL is an
interactive process.
For example, model checking itself may be specified (we say ``embedded'') in Higher Order Logic, \cite{kam:16b}.
Then the process of model checking can be modeled and its mechanisms be derived from first principles and, within
this embedding, application of model checking to application examples can be performed in Isabelle using its proof
support system. However, model checking actual application examples is not automated when using this embedded model
checking.

Clearly, this is a highly complex proof task and needs interaction to an extent that may make realistic application
examples infeasible. However, the advantages of a highly reliable process of model checking and its provably correct
application may make such efforts desirable for highly safety, security or privacy critical applications. In particular,
in human centric applications including complex infrastructures and policies, as for safety, security and privacy
critical applications \cite{kam:17b,kam:17c}, often the application domain poses challenges that also require
expressive logics in order to express multi-layered and multi-faceted aspects of the real world \cite{kam:19a}.
For example, human disposition, like in Insider threats for airplane security policies \cite{kk:21}, or protocols
including probabilities for quantum cryptography, e.g., \cite{kam:19c}.

In order to address the challenge of Higher Order Logic model checking, we have extended the existing model checking
theory \cite{kam:17a, kam:18b} provided for the Isabelle Insider and Infrastructure framework (IIIf) with an efficient
algorithm for model checking of attack trees. Despite the theoretical impossibility of
model checking HOL specifications, we provide within Isabelle a framework that allows automated (and provably correct)
extraction of model checking procedures.
This contribution is reported on in this paper. Necessary background is added
(Section \ref{sec:back}). The formalization of the model checking algorithm is described in Section \ref{sec:mc} and
illustrated on an application example (Section \ref{sec:app}). All Isabelle sources are available online.

\section{Background and Related Work}
\label{sec:back}
This section provides an overview of the current extension of the Isabelle Infrastructure
framework in relation to previous works and how it integrates the Refinement-Risk cycle
(Section \ref{sec:isa}). It also summarizes the formalization of the existing theories for Kripke structures 
and the temporal logic CTL (Section \ref{sec:kripke}), as well as the attack tree formalization
and Correctness and Completeness theorems (Section \ref{sec:at}). 
%Finally, Section \ref{sec:hcapp} presents the IoT Healthcare system -- the case study on which 
%the Refinement-Risk cycle is validated in this paper.

\subsection{Isabelle Infrastructure Framework}
\label{sec:isa}
Isabelle is a generic Higher Order Logic (HOL) proof assistant. Its generic
aspect allows the embedding of so-called object-logics as new theories
on top of HOL. There are sophisticated proof tactics available to support 
reasoning: simplification, first-order resolution, and special macros to support
arithmetic amongst others.
Object-logics are added to Isabelle using constant and type definitions
forming a so-called {\it conservative extension}. That is, no 
inconsistency can be introduced: new types are defined as 
subsets of existing types; properties are proved using a one-to-one 
relationship to the new type from properties of the existing type.
The use of HOL has the advantage that it enables expressing
even the most complex application scenarios, conditions, and logical
requirements. Isabelle enables the analysis of meta-theory, that is,
we can prove theorems {\it in} an object logic but also {\it about} it. 

This allows the building of telescope-like structures in which a meta-theory
at a lower level embeds a more concrete ``application'' at a higher level.
Properties are proved at each level.
Interactive proof is used to prove these 
properties but the meta-theory can be applied to immediately produce results.
%Figure \ref{fig:theorystruc} in Section \ref{sec:intro}
%gives an overview of the Isabelle Infrastructure 
%framework with its layers of object-logics -- each level below embeds the one
%above.
%\begin{figure}[h!]
%\begin{center}
%\includegraphics[scale=.4]{theory_structure}
%\end{center}
%%\vspace{-.5cm}
%\caption{Generic framework for infrastructures with model transformations.}
%\label{fig:theorystruc}
%\end{figure}

The Isabelle Insider and Infrastructure framework (IIIf) has been created initially for the modeling 
and analysis of Insider threats \cite{bikp:14}. Its use has been validated on the 
most well-known insider threat patterns identified by the CERT-Guide to Insider 
threats \cite{cmt:12}. It has been applied to Insider threats many times more recently being extended to
unintentional insiders \cite{kam:22b}.
The human centricity and logical foundation also are useful for Explainability of AI \cite{kam:22c, kam:22d}.
Moreover, this Isabelle framework has been successfully applied to realistic
case studies of insider attacks in airplane safety \cite{kk:16,kk:21} and on auction protocols
\cite{kkp:16,kkp:16a}.
These larger case studies as well as complementary work on the analysis
of Insider attacks on IoT infrastructures, e.g. \cite{knp:16},
have motivated the extension of the original framework by Kripke structures and 
temporal logic \cite{kam:16b} as well as a formalization of attack trees \cite{kam:18b}.
As a privacy relevant application, GDPR compliance verification has been demonstrated \cite{kam:18a}.
For security, the use of IIIf for interblockchain protocols is worth mentioning \cite{kn:20}.

The Refinement-Risk cycle (RR-cycle) \cite{kam:21a} conceptualizes the security engineering
process identified by these case studies. It allows repeated security improvements of the system
specification using identified security attacks. The RR-cycle formalizes a property preserving notion of
(often called horizontal) refinement combining it with the powerful locale mechanism of Isabelle
\cite{kpw:99,kam:00, kp:99} to support application example refinement and model checking. This mechanism is
now -- as reported in the current paper -- additionally extended by a process of ``vertical refinement''
by providing finite implementations of a system specification that enables fully automated model checking
for the attack tree analysis.
For initial specifications, the formal notion of Fusion Object-Z may be used as it enables to boot strap
a formal specification \cite{bk:03}.
One case study applies the step-by-step refinement of the RR-cycle to a security and privacy critical IoT
healthcare application. It exemplifies the refinement process interspersed with attack analysis using attack
trees to increase privacy by ultimately introducing a blockchain for access control \cite{kam:19a}.
It further integrates blockchain into the IoT Healthcare example \cite{kam:23b}.
Another application of the RR-cycle is the Corona-virus Warn App \cite{kl:20}. However, these first two application
of a systematic refinement do not yet consider the automation of model checking and the idea of a vertical
refinement as presented in this paper.

Concerning verification, an embedding of Extended Hierarchical Automata in Isabelle/HOL \cite{hk:01,hk:10} is
related to our work but the focus is on finding property preserving abstractions of the data within the
Statechart \cite{hk:05} to enable model checking. Even though model checking is also involved there, abstraction is
precisely the opposite of refinement.
To find a refinement, that is, design the refined specification and define the refinement relation is a
creative and ambiguous process; it reflects the development skill of a programmer.
An approach to derive refinement relations and refined specification for Object-Z specifications based on
heuristics is \cite{ks:04}.

\subsection{Kripke Structures and CTL}
\label{sec:kripke}
Kripke structures allow modeling state based systems and the temporal logic CTL (Computation Tree
Logic) \cite{cgp:99} enables analysis of properties under dynamic state changes.
A state transition relates snapshots of systems. These snapshots are the states of the Kripke structure. 
The temporal logic CTL then enables expressing security and privacy properties over the state transition
relation. 

In Isabelle, the system states and their transition relation are defined as a 
class called \texttt{state} containing an abstract constant \texttt{state\_transition}. 
It introduces the syntactic infix notation \texttt{I \ttrelI\, I'} to denote 
that system state \texttt{I} and \texttt{I'} are in this relation over an arbitrary 
(polymorphic) type $\ttsigma$. The operator \texttt{::} is a type judgement to 
coerce the type variable $\ttsigma$ into the class type. The arrow \texttt{\ttfun}
is the operator for functions on types and \texttt{bool} is the HOL inbuilt type
of truth values \texttt{true} and \texttt{false}.
\begin{ttbox}
 {\bf{class}} state =    
 {\bf{fixes}} state_transition :: (\ttsigma :: type) \ttfun \ttsigma \ttfun bool       
 ("_  \ttrelI _")
\end{ttbox}
The above class definition lifts Kripke structures and CTL to 
a general level.
The definition of the inductive relation is given by a set of specific rules
which are, however, %not necessary to define the notion of a Kripke structure
%and attack trees. They are 
part of an application like infrastructures (Section \ref{sec:app}).
Branching time temporal logic CTL %has been integrated as part of the
%Isabelle Insider framework \cite{kam:16b} built 
is defined in general over Kripke structures with arbitrary state transitions 
and can later be applied to suitable theories, like infrastructures. 
%and state transitions and applied to the Insider theory. We generalise
%this theory here using the above class definition and thereby lifting
%Kripke structures and CTL to a generic level.

Based on the generic state transition $\ttrelI$ of the type class \texttt{state},
the CTL-operators \texttt{\sf EX} and \texttt{\sf AX} express that property $f$ 
holds in some or all next states, respectively.
%\begin{ttbox}
%{\sf AX} \ttf \ttequiv \{ s. \{f0. s \ttrelI f0 \} \ttsubseteq \ttf \}
%{\sf EX} \ttf \ttequiv \{ s. \ttexists f0 \ttin \ttf. s \ttrelI f0 \}
%\end{ttbox}
The CTL formula \texttt{\sf AG} $f$ means that on all paths branching from 
a state $s$ the formula $f$ is always true (\texttt{\sf G} stands for `globally'). 
It can be defined using the Tarski fixpoint theory by applying the greatest 
fixpoint operator.
%\begin{ttbox}
% {\sf AG} \ttf \ttequiv gfp(\ttlam Z. \ttf \ttcap {\sf AX} Z)
%\end{ttbox}
In a similar way, the other CTL operators are defined. 
The formal Isabelle definition of what it means that formula 
$f$ holds in a Kripke structure \texttt{M} can be
stated as: the initial states of the Kripke structure \texttt{init M} 
need to be contained in the set of all states \texttt{states M} 
that imply $f$.
\begin{ttbox}
 M \ttvdash f \ttequiv  init M \ttsubseteq \{ s \ttin states M. s \ttin f \}
\end{ttbox}
In an application, the set of states of the Kripke structure is defined 
as the set of states reachable by the infrastructure state transition from 
some initial state, say \texttt{ex\_scenario}.
\begin{ttbox}
  ex_states \ttequiv \{ I. ex_scenario \ttrelIstar  I \}
\end{ttbox}
The relation \texttt{\ttrelIstar} is the reflexive transitive closure -- an operator
supplied by the Isabelle theory library -- applied to the relation \texttt{\ttrelI}.

The \texttt{Kripke} constructor combines the constituents initial state and state set.
% and state transition relation \texttt{\ttrelI}. %and labelling function \texttt{L}.
\begin{ttbox}
 ex_Kripke \ttequiv Kripke ex_states \{ex_scenario\} 
\end{ttbox}
%Properties in HOL are given as predicates. Alternatively, they can
%be seen a sets in a view which is often called {\it predicate transformer}
%semantics. 
%In HOL, sets are defined as predicates -- so these concepts coincide.
%Given some \texttt{property} -- a predicate over states (or equally a set of 
%states) --  we can thus then for example try to 
%prove that there is a path ({\sf E}) to a state in which the property 
%eventually holds (in the {\sf F}uture) by attempting the following proof 
%in Isabelle.
In Isabelle, the concept of sets and predicates coincide (more precisely they are isomorphic)
\footnote{In general, this is often referred to as {\it predicate transformer 
semantics.}}.
Thus a \texttt{property} is a predicate over states which is equal to a set of 
states. For example, we can then try to prove that there is a path ({\sf E}) 
to a state in which the property eventually holds (in the {\sf F}uture) by 
starting the following proof in Isabelle.
\begin{ttbox}
 ex_Kripke \ttvdash {\sf EF} property 
\end{ttbox}
Since \texttt{property} is a set of states, and the temporal operators
are predicate transformers, that is, transform sets of states to sets of states, 
the resulting {\sf EF} \texttt{property} is also a set of states -- and hence 
again a property.

\subsection{Attack Trees in Isabelle}
\label{sec:at}
\begin{figure}[h!]
\begin{center}
\includegraphics[scale=.25]{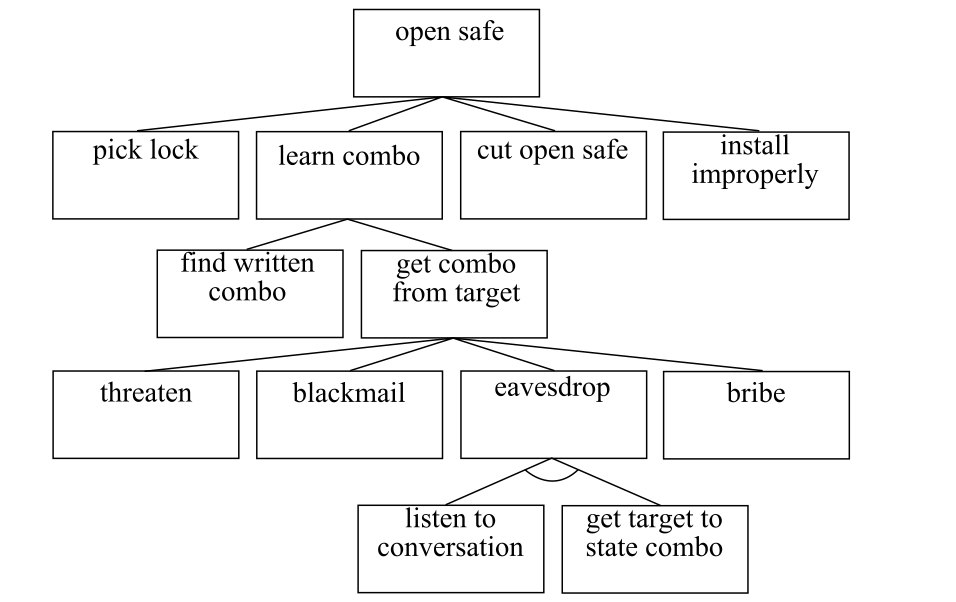}
\end{center}
\caption{Attack tree example illustrating disjunctive nodes for alternative 
attacks refining the attack ``open safe''. Near the leaves there is also a 
conjunctive node ``eavesdrop''.}
\label{fig:atex}
\end{figure}

Attack trees \cite{Schneier.102} are a graphical language for the analysis 
and quantification of attacks. If the root represents an attack, 
its children represent the sub-attacks. 
Leaf nodes are the basic attacks; other
nodes of attack trees represent sub-attacks.
%can be and-nodes and or-nodes that combine their 
%sub-trees either in a conjunctive or in a disjunctive manner. 
Sub-attacks can be alternatives for reaching the goal (disjunctive node) or 
they must all be completed to reach the goal (conjunctive node). 
%\TODO{Insert explanation to previous reviewer about parallev vs seq AND}
% Done see below at the point of the definition of the is_attack_tree predicate
Figure \ref{fig:atex} is an example of an attack tree taken from a textbook
\cite{Schneier.102} illustrating the attack of opening a safe.
Nodes can be adorned with attributes, for example costs of attacks
or probabilities which allows quantification of attacks (not used 
in the example in this paper but in \cite{kam:19c}).

The following datatype definition \texttt{attree} defines attack trees.
Isabelle allows recursive datatype definitions similar to the programming languages
Haskell or ML. A datatype is given by a ``\texttt{|}'' separated sequence of
possible cases each of which consists of a constructor name, the types of inputs to this
constructor, and optionally a pretty printing syntax definition.
The simplest case of an attack tree is a base attack.
The principal idea is that base attacks are defined by a pair of
state sets representing the initial states and the {\it attack property}
-- a set of states characterized by the fact that this property holds
for them. 
Attacks can also be combined as the conjunction or disjunction of other attacks. 
The operator $\oplus_\vee$ creates or-trees and $\oplus_\wedge$ creates and-trees.
And-attack trees $l \ttattand s$ and or-attack trees $l \ttattor s$ 
consist of a list of sub-attacks -- again attack trees. 
\begin{ttbox}
{\bf datatype} (\ttsigma :: state)attree = 
  BaseAttack (\ttsigma set)\tttimes(\ttsigma set) ("\ttcalN (_)") 
| AndAttack (\ttsigma attree)list (\ttsigma set)\tttimes(\ttsigma set) ("_ {\ttattand{(\_)}}")
| OrAttack  (\ttsigma attree)list (\ttsigma set)\tttimes(\ttsigma set) ("_ {\ttattor{(\_)}}")
\end{ttbox}
The attack goal is given by the pair of state sets on the right of the operator 
\texttt{\ttcalN}, $\oplus_\vee$ or $\oplus_\wedge$, respectively. A corresponding 
projection operator is defined as the function \texttt{attack}.

When we develop an attack tree, we proceed from an abstract attack, given
by an attack goal, by breaking it down into a series of sub-attacks. This
proceeding corresponds to a process of {\it refinement}. 
The attack tree calculus \cite{kam:18b} provides a notion of attack tree refinement
elegantly expressed as the infix operator $\sqsubseteq$. 
Note that this refinement is different from the notion of system refinement
%that will be presented later in this paper.
that builds the core mechanism of the Refinement-Risk cycle (RR-cycle) \cite{kam:21a}
allowing repeated security improvements of the system specification using identified
security attacks. 

\begin{figure*}
\begin{center}
\includegraphics[scale=.3]{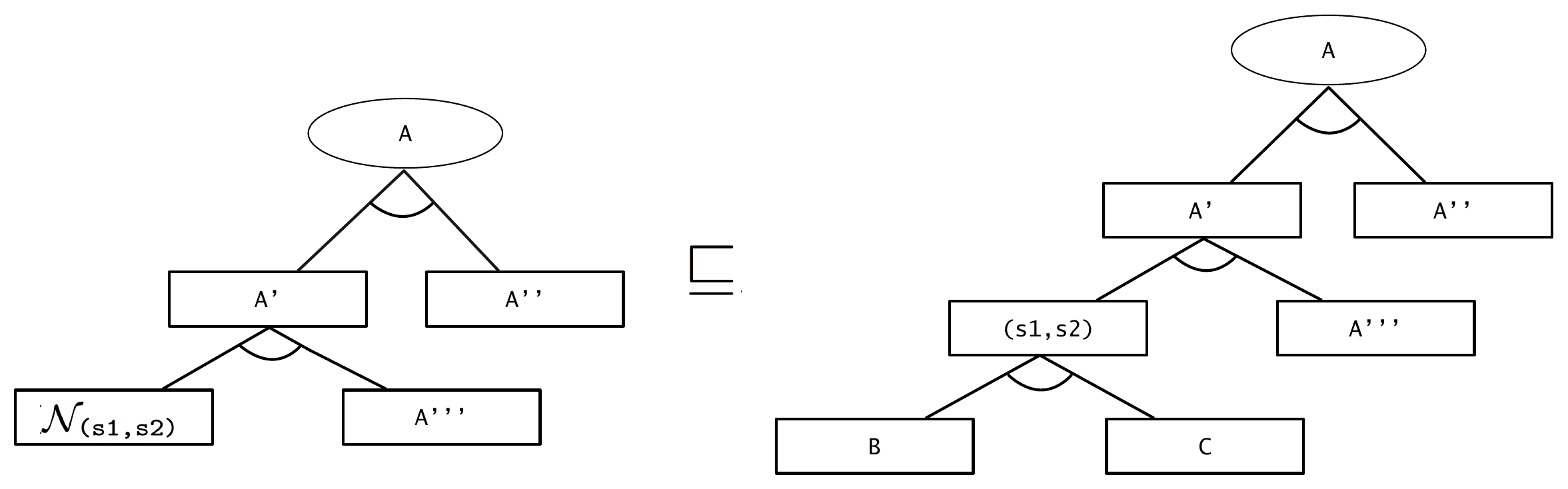}
\end{center}
%\vspace{-.5cm}
\caption{Attack tree example illustrating refinement of an and-subtree.}
\label{fig:ref}
\end{figure*}

The intuition of developing an attack tree by refinement from the root to the leaves
is illustrated  in Figure \ref{fig:ref} (the formal definition is in \cite{kam:18b}).
The example attack tree on the left side has a leaf that is expanded by the refinement 
into an and-attack with two steps.
Refinement of attack trees defines the stepwise process of expanding abstract
attacks into more elaborate attacks only syntactically. There is no guarantee
that the refined attack is possible if the abstract one is, nor vice-versa.
The attack tree calculus \cite{kam:18b} formalizes the semantics of attack trees
on Kripke structures and CTL enabling rigorous judgement whether such
syntactic refinements represent possible attacks.

A valid attack, intuitively, is one which is fully refined into fine-grained
attacks that are feasible in a model. The general model provided is
a Kripke structure, i.e., a set of states and a generic state transition.
Thus, feasible steps in the model are single steps of the state transition.
They are called valid base attacks.
The composition of sequences of valid base attacks into and-attacks yields
again valid attacks if the base attacks line up with respect to the states
in the state transition. If there are different valid attacks for the same
attack goal starting from the same initial state set, these can be 
summarized in an or-attack. The formal definition \cite{kam:18b} is given in the table
in Figure \ref{fig:isatttree}.
\begin{figure*}
\begin{ttbox}
{\bf fun} is_attack_tree :: [(\ttsigma :: state) attree] \ttfun bool  ("\ttvdash_") 
{\bf where} 
  att_base:  \ttvdash \ttcalN{s} = \ttforall x \ttin fst s. \ttexists y \ttin snd s. x  \ttrelI y  
| att_and: \ttvdash (As :: (\ttsigma::state attree list)) \ttattand{s} = 
           case As of
             [] \ttfun (fst s \ttsubseteq snd s)
           |  [a] \ttfun \ttvdash a \ttand attack a = s 
           |  a \# l \ttfun \ttvdash a \ttand fst(attack a) = fst s 
                         \ttand \ttvdash l \ttattand{\texttt{(snd(attack a),snd(s))}} 
| att_or: \ttvdash (As :: (\ttsigma::state attree list)) \ttattor{s} = 
          case As of 
             [] \ttfun (fst s \ttsubseteq snd s) 
          | [a] \ttfun \ttvdash a \ttand fst(attack a) \ttsupseteq fst s \ttand snd(attack a) \ttsubseteq snd s
          | a \# l \ttfun \ttvdash a \ttand fst(attack a) \ttsubseteq fst s \ttand snd(attack a) \ttsubseteq snd s
                       \ttand \ttvdash l \ttattor{\texttt{(fst s - fst(attack a),snd s)}}
\end{ttbox}
\caption{Definition of attack tree validity as one recursive predicate.}\label{fig:isatttree}
\end{figure*}
The semantics of attack trees is described by this one recursive function. Since the definition
can be given as a recursive function, Isabelle code generation is applicable: an executable
decision procedure for attack tree validity can be automatically generated in various programming 
languages, for example, ML, Haskell, or Scala.

Adequacy of the semantics is proved in \cite{kam:18b} by proving correctness and completeness.
The following correctness theorem shows that
if \texttt{A} is a valid attack on property \texttt{s} starting from
initial states described by \texttt{I}, then from all states in \texttt{I} 
there is a path to the set of states fulfilling \texttt{s} in the 
corresponding Kripke structure.
\begin{ttbox}
{\bf theorem} AT_EF: \ttvdash A :: (\ttsigma :: state) attree) \ttImp 
 (I, s) = attack A \ttImp 
 Kripke \{t . \ttexists i \ttin I. i \ttrelI^* t\} I \ttvdash {\sf EF} s
\end{ttbox}
The inverse direction of theorem \texttt{AT\_EF} is a completeness
theorem: if states described by predicate \texttt{s} can be reached from a 
finite nonempty set of initial states \texttt{I} in 
a Kripke structure, then there exists a valid attack tree for the attack 
\texttt{(I,s)}.
\begin{ttbox} 
{\bf theorem} Completeness: I \ttneq \{\} \ttImp finite I \ttImp
 Kripke \{t . \ttexists i \ttin I. i \ttrelI^* t\} I \ttvdash {\sf EF} s \ttImp
 \ttexists A :: (\ttsigma::state)attree. \ttvdash A \ttand (I, s) = attack A 
\end{ttbox}
Correctness and Completeness are proved in Isabelle \cite{kam:18b, kam:18smc}.
They are not just necessary proofs on the attack tree semantics but the theorems
allow easy transformation of properties between the embedded notions
of attack tree validity $\vdash$ and CTL formulas like {\sf EF}.
The relationship between these notions can be applied to case studies. That is, if we 
apply attack tree refinement to spell out an abstract attack tree for attack \texttt{s} 
into a valid attack sequence, we can apply theorem \texttt{AT\_EF} and can immediately 
infer that {\sf EF} \texttt{s} holds. Vice versa, the theorem Completeness can be applied
to directly infer the existence of an attack tree from the former.

\section{Implementing Model Checking in Isabelle}
\label{sec:mc}
As we have seen, model checking is fully formalized in the Isabelle theory and adequacy to attack trees is proved.
In order to make the actual process of model checking efficient, executable definitions can be devised
in Isabelle and code into programming languages can be automatically generated from them. However, we adopt a
strategy different to the obvious straightforward way. It would seem more straightforward to directly use the
formalized Tarski theory to provide executable code for the fixpoint combinators used as the definition of the
CTL operators. In fact, Tarski's theory in Isabelle proves lemmas for monotone fixpoint combinators as finite
computation sequences.
These lemmas  are proved in Isabelle permitting the computation of the fixpoints as finite iterations starting
from either the empty set of states (False) or the set of all states (True). Instead of implementing executable
functions for the computation of such {\it general fixpoint combinators}, that is, a generic algorithm for the
implmentation of all CTL operators, we decided to be more pragmatic and focus on the main application at hand:
{\it security and attack trees}. We use the above summarized previous work on formalizing
attack trees, Kripke structures and CTL to provide model checking for attack trees directly. However, because of
the theory proved in our framework, we can show that our model checking procedure indeed solves the theoretical
model checking problem for security attacks.

Our approach is conceived by the following contributions.
\begin{itemize}
\item We provide an efficient state transition function for the state transiton relation $\ttrelI$.
  Instead of sets, functions and relations, the efficient state transition uses lists to represent sets
  and simulates these mathematical objects including their functions and relations.
  There is a mapping \texttt{EsetI} from every efficient infrastructure state (type \texttt{einfrastructure} over
  lists) to infrastructures.
  \begin{ttbox}
  \ttforall I :: infrastructure. finite\_Infrastructure I
  \ttImp (\ttexists E:: einfrastructure. EsetI E = I)
  \end{ttbox}
  
\item We further define this state transition function \texttt{st\_e} by implementing a computable version of
  the state transition relation as a function over the efficient infrastructure type \texttt{einfrastructure}.
  We prove that this efficient state transition \texttt{st\_e} corresponds to the state transition relation
  $\ttrelI$ for finite infrastructures. That is, we prove the correctness of \texttt{st\_e} with respect to 
  \texttt{EsetI}: the latter preserves the efficient state transition \texttt{st\_e}.
  \begin{ttbox}
   st\_e G h P D F = Some G'
   \ttImp EsetI (Einfrastructure G P D F) \ttrelI
      EsetI (Einfrastructure G' P D F)
  \end{ttbox}  
\item Additionally, for every finite infrastructure state transition $\ttrelI$ there is a corresponding
  \texttt{st\_e} transition between pairs of their efficient states, i.e., we prove completeness.
  That is, for any finite infrastructure state \texttt{s'} reachable from an initial state
  \texttt{I}, that is, in the transitive closure \texttt{\ttrelIstar} of the state transition relation, there is
  a sequence of inputs \texttt{hs} that takes an efficient representation \texttt{e} of the initial state \texttt{I}
  to a representation \texttt{e'} of \texttt{s'}.
\begin{ttbox}
I \ttrelIstar s' \ttImp finite_Infrastructure I \ttImp
EsetI (Einfrastructure e P D F) = I \ttImp
\ttexists hs e'.
  fold\_o (\ttlam x h. st\_e x h P D F) e hs = Some e' \ttand
  EsetI (Einfrastructure e' P D F) = s'
  \end{ttbox}  

\item We thus conclude that a finite number of efficient state transitions leading from $s_e$ to $s'_e$
  corresponds to the corresponding infrastructure states $s, s'$ being in the reflexive transitive closure
  of the state transition relation $\ttrelIstar$.
  From this we can directly conclude -- using the attack tree theory presented in the previous section --
  that ${\sf EF} s'$. Vice versa, if $s$ is an initial state and $s'$ belongs to a predicate representing
  an attack, there exists an attack path leading to the attack and thus an attack tree.
\item Finally, since the efficient state transition relation \texttt{st\_e} is executable, we can implement a
  process of collecting all reachable states by repeatedly applying \texttt{st\_e} to initial states.
  This collection function \texttt{st\_et} is executed with a bound $n$. We prove that if a state is contained in
  the resulting state collection then it is reachable.
\item Model checking can now be implemented simply by intersecting reachable states collected by
  \texttt{st\_et} and attack states. The function \texttt{st\_et} is computable with a bound $n$. Since we presume
  that the set of states is finite, we can prove that there exists an $n$ such that \texttt{st\_et} reaches a fixpoint.
  The additional assumptions can naturally be contained in a locale to support a structured and systematic vertical
  refinement.
  From the executable definition of the executable functions \texttt{st\_et} and intersection in Isabelle, we can
  automatically generate code and use it on our case study.
\end{itemize}

\section{Application Example IoT Healthcare System}
\label{sec:app}

\begin{figure}[h]
\begin{center}
\includegraphics[scale=.19]{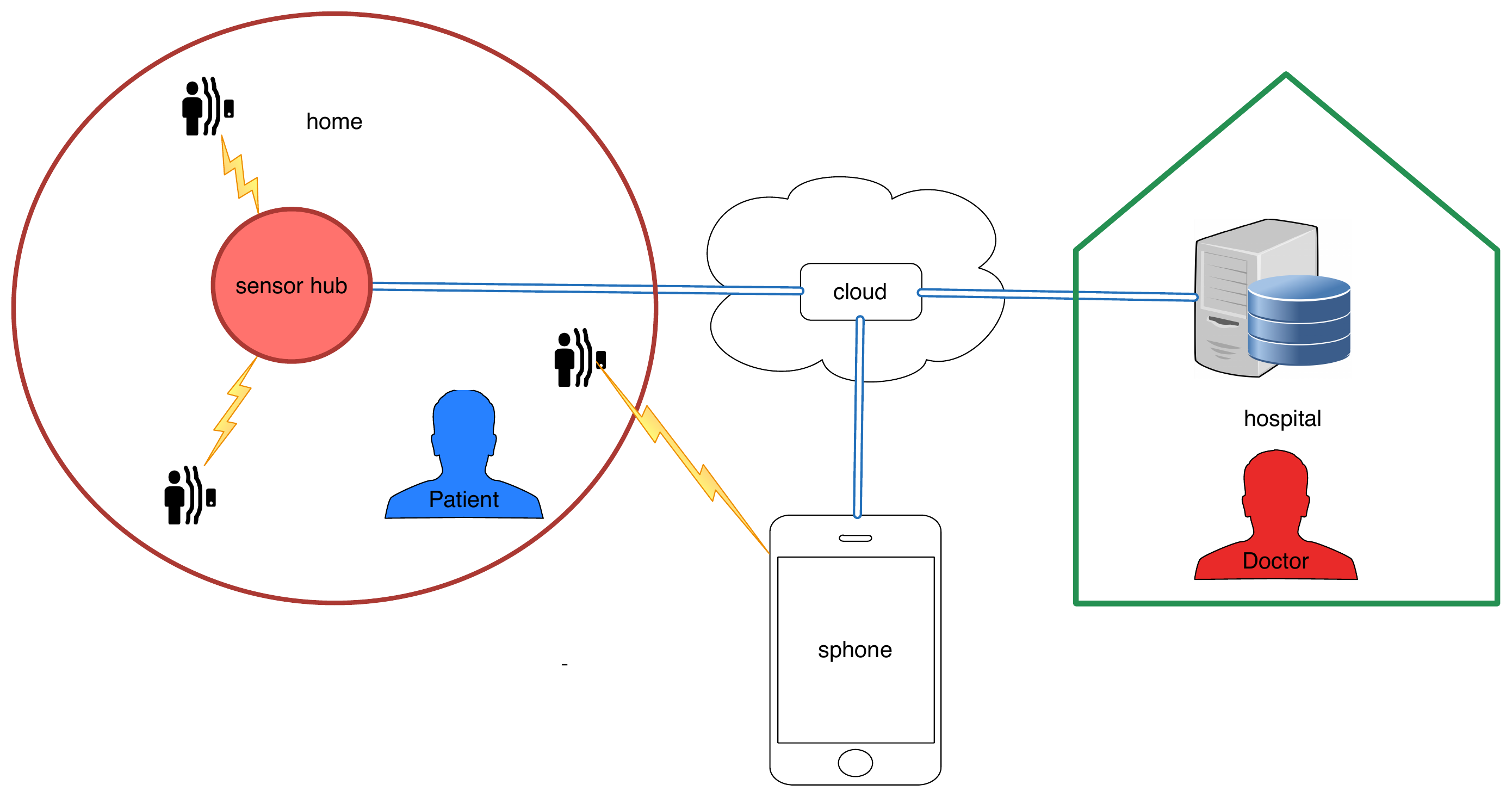}
\caption{IoT healthcare monitoring system for SUCCESS project \cite{suc:16}}\label{fig:iot}
\end{center}
%\vspace{-.5cm}
\end{figure}

%\subsection{IIIf: Infrastructures, Actors, and Policies}
%\label{sec:infra}
%\subsection{Edge Computing: IoT Healthcare System}
%\label{sec:hcapp}
% Describe the model of the IoT Healthcare system
%\subsection{SUCCESS Security and Privacy for IoT Healtcare}
%\label{sec:success}
% short summary of project and goals - could be part of some backgrond section
% or part of the intro to case study
%First, we introduce the application domain of transparent security and privacy of
The example of an IoT healthcare systems is from the CHIST-ERA project SUCCESS \cite{suc:16}
on monitoring Alzheimer's patients. % (Section \ref{sec:success}). 
Figure \ref{fig:iot} illustrates the system architecture where data collected by sensors 
in the home or via a smartphone helps monitoring bio markers of the patient. The data 
collection is in a cloud based server to enable hospitals (or scientific institutions) 
to access the data which is controlled via the smartphone.
It is a typical edge network application: the smartphone and the sensor hub in the 
home are typical edge devices that are capable of doing processing data without uploading
to the cloud server.

For the illustration of the model checking theory and its implementation as an executable attack tree
detection algorithm, we follow \cite{kam:23b} detecting privacy attacks in this scenario.

As the requirement specification, the following global policy expresses data privacy.
\begin{ttbox}
global_policy I a \ttequiv
  \ttforall l \ttin nodes(graphI I). \ttforall l' \ttin nodes(graphI I).
  \ttforall d:: data. \ttforall lb:: dlm. \ttforall lb':: dlm.
  (lb, d) \ttin lgra(graphI I l) \ttimp
  (lb', d) \ttin lgra(graphI I l') \ttimp
  lb = lb'
\end{ttbox}  
It expresses privacy by saying that different occurrences of the same data in the system must have
the same labels. Trying to prove this global policy, we fail.
The reason for this is exhibited by the following attack which we find by the automated attack tree model
checking analysis presented in this paper. After detection, the process outlined in the previous section
allows using the CTL semantics of attack trees in IIIf \cite{kam:18b}. The attack detected by the model checking
algorithm corresponds thus to an {\sf EF} property showing that there exists a path ({\sf E}) on which eventually
({\sf F}) Eve can put data on the cloud.
\begin{ttbox}
 hc_KripkeF \ttvdash
 {\sf EF} {I. enables I cloud (Actor ''Eve'') put}
\end{ttbox}
Even though the specification of the IoT system that we have specified uses data labeling and imposes 
privacy preserving functions, this attack is possible.
Eve can learn the data by other means than using the privacy preserving functions provided by the system.
Then she can use the action \texttt{put} to enter that learned private data as new data to the system but
labelled as her own data. Although abstract, this attack shows that attacks coming from outside the system
model may be detected by the system. Even though this does not explicitly model covert channels and implicit
flows, it allows to detect their desastruous effects.

\section{Conclusions}
We have shown in this paper that we can transform the model checking formalised in the IIIf into an
executable model checking algorithm that detects attacks corresponding to attack trees. The derivation of
this model checking algorithm corresponds to vertical refinement.
We have sketched how this model checking algorithm can be applied to an IoT healthcare example and thus
complements the RR-cycle.

%% Bibliography
%\bibliographystyle{abbrv}
%\bibliography{biblio}


%% Appendix
%\appendix
%\section{Appendix}
%
%Text of appendix \ldots

\end{document}